\documentclass[10pt,letterpaper]{article}
\usepackage{opex3}
\usepackage{color}
\usepackage{braket}
\usepackage{hyperref }
\usepackage{graphicx}
\usepackage{textcomp}
\usepackage{epstopdf}
\usepackage{amsmath}
\usepackage{amssymb}

\begin{document}

\title{Image formation properties and inverse imaging problem in aperture based scanning near field optical microscopy}
\author{S. Schmidt$^{1,2,*}$, A. E. Klein $^{2}$, T. Paul $^{1}$, H. Gross$^{1,2}$, S. Diziain $ ^{2} $, M. Steinert $ ^{2} $, A. C. Assafrao $ ^{3} $, T. Pertsch $ ^{2} $, H. P. Urbach $ ^{3} $ and C. Rockstuhl $ ^{4,5} $  }

\address{$^1$Fraunhofer Institute of Applied Optics and Precision Engineering, 07745 Jena, Germany\\
$^2$Friedrich-Schiller-Universit\"at Jena, Institute of Applied Physics, Abbe Center of Photonics, 07743 Jena, Germany\\
$ ^3 $ Optics Research Group, Department of Imaging Science and Technology, Delft University of
	Technology,  2628 CJ Delft, The Netherlands\\
	$ ^4 $ Institute of Theoretical Solid State Physics, Karlsruhe Institute of Technology, 76131 Karlsruhe, Germany\\
	$ ^5 $ Institute of Nanotechnology, Karlsruhe Institute of Technology, 76021 Karlsruhe, Germany}

\email{$^*$soeren.schmidt@iof.fraunhofer.de} 

\date{\today}
\begin{abstract}
Aperture based scanning near field optical microscopes are important instruments to study light at the nanoscale and to understand the optical functionality of  photonic nanostructures. In general, a detected image is affected by both, the transverse electric and magnetic field components of light. The discrimination of the individual field components is challenging, as these four field components are contained within two signals in the case of a polarization-resolved measurement. Here, we develop a methodology to solve the inverse imaging problem and to retrieve the vectorial field components from polarization- and phase-resolved measurements. Our methodology relies on the discussion of the image formation process in aperture based scanning near field optical microscopes. On this basis, we are also able to explain how the relative contributions of the electric and magnetic field components within detected images depend on the probe geometry, its material composition, and the illumination wavelength. This allows to design probes that are dominantly sensitive either to the electric or magnetic field components of light.
\end{abstract}
 \ocis{(180.4243) Near-field microscopy; (350.4238) Nanophotonics and photonic crystals;(250.5403) Plasmonics;(050.6624) Subwavelength structures } 
 \bibliography{Literatur}

\begin{thebibliography}{10}
\newcommand{\enquote}[1]{``#1''}

\bibitem{schnell2010amplitude}
M.~Schnell, A.~Garcia-Etxarri, A.~J. Huber, K.~B. Crozier, A.~Borisov,
  J.~Aizpurua, and R.~Hillenbrand, \enquote{Amplitude- and phase-resolved
  near-field mapping of infrared antenna modes by transmission-mode
  scattering-type near-field microscopyâ€ ,} The Journal of Physical
  Chemistry C \textbf{114}, 7341--7345 (2010).

\bibitem{ez2}
M.~Esslinger, J.~Dorfm\"uller, W.~Khunsin, R.~Vogelgesang, and K.~Kern,
  \enquote{Background-free imaging of plasmonic structures with cross-polarized
  apertureless scanning near-field optical microscopy,} Review of Scientific
  Instruments \textbf{83}, 033704 (2012).

\bibitem{ez1}
R.~Esteban, R.~Vogelgesang, J.~Dorfm\"uller, A.~Dmitriev, C.~Rockstuhl,
  C.~Etrich, and K.~Kern, \enquote{Direct near-field optical imaging of higher
  order plasmonic resonances,} Nano Letters \textbf{8}, 3155--3159 (2008).

\bibitem{severine}
S.~Diziain, D.~Barchiesi, T.~Grosges, and P.-M. Adam, \enquote{Recovering of
  the apertureless scanning near-field optical microscopy signal through a
  lock-in detection,} Applied Physics B \textbf{84}, 233--238 (2006).

\bibitem{Feber2013}
B.~le~Feber, N.~Rotenberg, D.~M. Beggs, and L.~Kuipers, \enquote{Simultaneous
  measurement of nanoscale electric and magnetic optical fields,} Nature Photon
  \textbf{8}, 43--46 (2013).

\bibitem{Feber20142}
B.~le~Feber, N.~Rotenberg, D.~van Oosten, and L.~Kuipers, \enquote{Modal
  symmetries at the nanoscale: a route toward a complete vectorial near-field
  mapping,} Optics Letters \textbf{39}, 2802 (2014).

\bibitem{Rotenberg2014}
N.~Rotenberg and L.~Kuipers, \enquote{Mapping nanoscale light fields,} Nature
  Photon \textbf{8}, 919--926 (2014).

\bibitem{walford2002}
J.~N. Walford, J.-A. Porto, R.~Carminati, and J.-J. Greffet, \enquote{Theory of
  near-field magneto-optical imaging,} JOSA A \textbf{19}, 572--583 (2002).

\bibitem{Porto2000}
J.~A. Porto, R.~Carminati, and J.-J. Greffet, \enquote{Theory of
  electromagnetic field imaging and spectroscopy in scanning near-field optical
  microscopy,} Journal of Applied Physics \textbf{88}, 4845 (2000).

\bibitem{dev1}
E.~Devaux, A.~Dereux, E.~Bourillot, Y.~Weeber, J.C.and~Lacroute, J.~Goudonnet,
  and C.~Girard, \enquote{Local detection of the optical magnetic field in the
  near zone of dielectric samples,} Phys. Rev. B \textbf{62}, 10504 (2000).

\bibitem{dev2}
A.~Dereux, E.~Devaux, J.~C. Weeber, J.~P. Goudonnet, and C.~Girard,
  \enquote{Direct interpretation of near-field optical images,} Journal of
  Microscopy \textbf{202}, 320--331 (2001).

\bibitem{Angelaneu}
A.~E. Klein, N.~Janunts, M.~Steinert, A.~T\"unnermann, and T.~Pertsch,
  \enquote{Polarization-resolved near-field mapping of plasmonic aperture
  emission by a dual-snom system,} Nano Letters  (2014).

\bibitem{ang1}
A.~E. Klein, A.~Minovich, M.~Steinert, N.~Janunts, A.~T\"unnermann, and
  T.~Pertsch, \enquote{Controlling plasmonic hot spots by interfering airy
  beams,} Opt. Lett. \textbf{37}, 3402--3404 (2012).

\bibitem{bur2}
M.~Burresi, D.~van Oosten, T.~Kampfrath, H.~Schoenmaker, R.~Heideman,
  A.~Leinse, and L.~Kuipers, \enquote{Probing the magnetic field of light at
  optical frequencies,} Science \textbf{326}, 550--553 (2009).

\bibitem{ang2}
A.~Minovich, A.~E. Klein, N.~Janunts, and T.~Pertsch, \enquote{Generation and
  near-field imaging of airy surface plasmons,} Phys. Rev. Lett. \textbf{107},
  116802 (2011).

\bibitem{bou1}
A.~Bouhelier, F.~Ignatovich, A.~Bruyant, C.~Huang, G.~Colas~des Francs, and
  J.-C. Weeber, \enquote{Surface plasmon interference excited by tightly
  focused laser beams,} Optics Letters \textbf{32}, 2535--2537 (2007).

\bibitem{bur1}
M.~Burresi, A.~Engelen, R.~Opheij, D.~van Oosten, D.~Mori, T.~Baba, and
  L.~Kuipers, \enquote{Observation of polarization singularities at the
  nanoscale,} Phys. Rev. Lett. \textbf{102}, 2480--2483 (2009).

\bibitem{bur4}
M.~Burresi, T.~Kampfrath, D.~van Oosten, J.~C. Prangsma, B.~S. Song, S.~Noda,
  and L.~Kuipers, \enquote{Magnetic light-matter interactions in a photonic
  crystal nanocavity,} Phys. Rev. Lett. \textbf{105}, 123901 (2010).

\bibitem{uebel2012}
P.~Uebel, M.~A. Schmidt, H.~W. Lee, and P.~S. Russell,
  \enquote{Polarisation-resolved near-field mapping of a coupled gold nanowire
  array,} Optics Express \textbf{20}, 28409 (2012).

\bibitem{Denkova2013}
D.~Denkova, N.~Verellen, A.~V. Silhanek, V.~K. Valev, P.~V. Dorpe, and V.~V.
  Moshchalkov, \enquote{Mapping magnetic near-field distributions of plasmonic
  nanoantennas,} ACS Nano \textbf{7}, 3168--3176 (2013).

\bibitem{Denkova2014}
D.~Denkova, N.~Verellen, A.~V. Silhanek, P.~Van~Dorpe, and V.~V. Moshchalkov,
  \enquote{Near-field aperture-probe as a magnetic dipole source and optical
  magnetic field detector,} arXiv preprint arXiv:1406.7827  (2014).

\bibitem{koh1}
D.~C. Kohlgraf-Owens, S.~Sukhov, and A.~Dogariu, \enquote{Discrimination of
  field components in optical probe microscopy,} Optics Letters \textbf{37},
  3606--3608 (2012).

\bibitem{kihm1}
H.~W. Kihm, S.~M. Koo, Q.~H. Kim, K.~Bao, J.~E. Kihm, W.~S. Bak, S.~H. Eah,
  C.~Lienau, H.~Kim, P.~Nordlander \emph{et~al.}, \enquote{Bethe-hole
  polarization analyser for the magnetic vector of light,} Nature
  Communications \textbf{2}, 451 (2011).

\bibitem{Kihm2}
H.~W. Kihm, J.~Kim, S.~Koo, J.~Ahn, K.~Ahn, K.~Lee, and N.~Park,
  \enquote{Optical magnetic field mapping using a subwavelength aperture,}
  Optics Express \textbf{21}, 5625--5633 (2013).

\bibitem{het2}
A.~Nesci, R.~D\"{a}ndliker, and H.~P. Herzig, \enquote{Quantitative amplitude
  and phase measurement by use of a heterodyne scanning near-field optical
  microscope,} Opt. Lett. \textbf{26}, 208--210 (2001).

\bibitem{bur3}
M.~Burresi, D.~Diessel, D.~van Oosten, S.~Linden, M.~Wegener, and L.~Kuipers,
  \enquote{Negative-index metamaterials: Looking into the unit cell,} Nano
  Letters \textbf{10}, 2480--2483 (2010).

\bibitem{grosjean2010}
T.~Grosjean, I.~Ibrahim, M.~Suarez, G.~Burr, M.~Mivelle, and D.~Charraut,
  \enquote{Full vectorial imaging of electromagnetic light at subwavelength
  scale,} Optics express \textbf{18}, 5809--5824 (2010).

\bibitem{caselli2015}
N.~Caselli, F.~La~China, W.~Bao, F.~Riboli, A.~Gerardino, L.~Li, E.~H.
  Linfield, F.~Pagliano, A.~Fiore, P.~J. Schuck \emph{et~al.},
  \enquote{Deep-subwavelength imaging of both electric and magnetic localized
  optical fields by plasmonic campanile nanoantenna,} Scientific reports
  \textbf{5} (2015).

\bibitem{gre1}
J.~J. Greffet and R.~Carminati, \enquote{Image formation in near-field optics,}
  Progress in surface science \textbf{56}, 133--237 (1997).

\bibitem{VanLabeke1993}
D.~Van~Labeke and D.~Barchiesi, \enquote{Probes for scanning tunneling optical
  microscopy: a theoretical comparison,} Journal of the Optical Society of
  America A \textbf{10}, 2193 (1993).

\bibitem{Hecht2000}
B.~Hecht, B.~Sick, U.~P. Wild, V.~Deckert, R.~Zenobi, O.~J.~F. Martin, and
  D.~W. Pohl, \enquote{Scanning near-field optical microscopy with aperture
  probes: Fundamentals and applications,} The Journal of Chemical Physics
  \textbf{112}, 7761 (2000).

\bibitem{novotny1994}
L.~Novotny and C.~Hafner, \enquote{Light propagation in a cylindrical waveguide
  with a complex, metallic, dielectric function,} Phys. Rev. E \textbf{50},
  4094--4106 (1994).

\bibitem{novotny}
L.~Novotny and B.~Hecht, \enquote{Principles of nano-optics,} Cambridge
  University Press  (2006).

\bibitem{sny}
A.~W. Snyder and J.~Love, \enquote{Optical waveguide theory,} Springer  (1983).

\bibitem{bakker2015}
R.~M. Bakker, D.~Permyakov, Y.~F. Yu, D.~Markovich, R.~Paniagua-Dom{\'\i}nguez,
  L.~Gonzaga, A.~Samusev, Y.~Kivshar, B.~Luk’yanchuk, and A.~I. Kuznetsov,
  \enquote{Magnetic and electric hotspots with silicon nanodimers,} Nano
  Letters \textbf{15}, 2137--2142 (2015).

\bibitem{sinev2015}
I.~S. Sinev, P.~M. Voroshilov, I.~S. Mukhin, A.~I. Denisyuk, M.~E. Guzhva,
  A.~K. Samusev, P.~A. Belov, and C.~R. Simovski, \enquote{Demonstration of
  unusual nanoantenna array modes through direct reconstruction of the
  near-field signal,} Nanoscale \textbf{7}, 765--770 (2015).

\bibitem{gross2005handbook}
B.~D{\"o}rband, H.~Gross, and H.~M{\"u}ller, \enquote{Handbook of optical
  systems, metrology of optical components and systems,} Wiley-Blackwell
  (2012).

\bibitem{singer2006handbook}
H.~Gross, W.~Singer, and M.~Totzeck, \enquote{Handbook of optical systems,
  physical image formation,} John Wiley \& Sons \textbf{2} (2006).

\bibitem{Esslinger2012}
M.~Esslinger and R.~Vogelgesang, \enquote{Reciprocity theory of apertureless
  scanning near-field optical microscopy with point-dipole probes,} ACS Nano
  \textbf{6} (2012).

\bibitem{Rockstuhl2008}
C.~Rockstuhl, T.~Zentgraf, T.~P. Meyrath, H.~Giessen, and F.~Lederer,
  \enquote{Resonances in complementary metamaterials and nanoapertures,} Optics
  Express \textbf{16}, 2080 (2008).

\bibitem{Abajo2007}
F.~J. Garcia~de Abajo, \enquote{Colloquium: Light scattering by particle and
  hole arrays,} Rev. Mod. Phys. \textbf{79}, 1267--1290 (2007).

\bibitem{dorn2003}
R.~Dorn, S.~Quabis, and G.~Leuchs, \enquote{Sharper focus for a radially
  polarized light beam,} Physical review letters \textbf{91}, 233901 (2003).

\bibitem{novotny2001}
L.~Novotny, M.~Beversluis, K.~Youngworth, and T.~Brown, \enquote{Longitudinal
  field modes probed by single molecules,} Physical review letters \textbf{86},
  5251 (2001).

\bibitem{gro12}
L.~Rabiner, R.~Schafer, and C.~Rader, \enquote{The chirp z-transform
  algorithm,} Audio and Electroacoustics, IEEE Transactions on \textbf{17},
  86--92 (1969).

\bibitem{gro2}
S.~Roose, B.~Brichau, and E.~Stijns, \enquote{An efficient interpolation
  algorithm for {Fourier} and diffractive optics,} Optics communications
  \textbf{97}, 312--318 (1992).

\bibitem{gro3}
J.~Bakx, \enquote{Efficient computation of optical disk readout by use of the
  chirp z transform,} Applied Optics \textbf{41}, 4897--4903 (2002).

\bibitem{lal}
W.~Smigaj, P.~Lalanne, J.~Yang, T.~Paul, C.~Rockstuhl, and F.~Lederer,
  \enquote{Closed-form expression for the scattering coefficients at an
  interface between two periodic media,} Applied Physics Letters \textbf{98},
  111107 (2011).

\end{thebibliography}
 \section{Introduction}
 In nano-optics, the domain of research that deals with the interaction of light with objects having a critical length scale in the order of a few up to hundreds of nanometers, the optical near field is a key quantity.  Only in the near field the fraction of the angular spectrum that is associated with evanescent waves has a notable amplitude. These evanescent field components carry the high spatial frequency information of the underlying electro-magnetic field. Experimental access to the near fields, therefore, is often important to understand the interplay between photonic nanostructures and an incident light field. This triggered the development of scanning near field optical microscopes (SNOM). These instruments rely on the perturbation of the near field by a sharp tip. Roughly spoken, in this scattering process evanescent waves are converted to propagating waves that can be detected by classical instruments in the far field. This basic idea led to different instrumental implementations.  
 
 In an apertureless (also called scattering) SNOM, a tip with an apex size of several nanometers scatters the light from the near into the far field \cite{schnell2010amplitude,ez2,ez1,severine}.
 Alternatively, more compact devices that base on an aperture version of the SNOM were developed. There, a tapered waveguide is covered with a metallic film. A small opening at the apex  locally collects the light. For such aperture SNOM, which we study in this contribution, it was predicted and proven that detected signals are influenced by both the transverse electric and magnetic field components \cite{Feber2013,Feber20142,Rotenberg2014,walford2002,Porto2000}. Depending on the measurement configuration, a detected signal can be influenced dominantly by either the transverse electric, magnetic, or both field components \cite{dev1,dev2,Angelaneu,ang1,bur2,ang2,bou1,bur1,bur4,uebel2012,Feber2013,Denkova2013,Denkova2014,koh1,kihm1,Kihm2,bur2,Feber20142,Rotenberg2014}. 
 
 Here, we discuss in detail the properties that lead to the dominant detection of either the electric or the magnetic field components. We prove that the geometry and the material composition of the probe at the measurement wavelength strongly affect which field-component is eventually measured. These insights provide indications for aperture-SNOM probes that measure dominantly the electric or the magnetic field components. Exemplarily, we study a specific probe coated by either gold or aluminum that possess a dominant sensitivity to either the magnetic or electric field components, respectively.
 
 Based on the discussion of the image formation process, we outline a methodology to solve the inverse imaging problem. By relying on a polarization- and phase-resolved measurement, being available with the current state-of-the-art \cite{het2,Feber2013,Feber20142,bur1,bur2,bur3,bur4}, two distinct signals are gathered. These two signals correspond to the complex amplitudes of two orthogonally polarised modes of the fiber that transmit the signal to the far field.
 These two complex signals, in general, are influenced by four complex quantities, namely the transverse in-plane electric and magnetic field components. The coupling of these field components in the process of collecting the signal by the probe and their simultaneous contribution to the two measurement signals makes it challenging to directly compare measured images in the two signals to any of the components of the field to be measured. 
 
 It was proven already that with suitable \textit{a priori} knowledge, the coupling can be resolved and the information about the investigated field components in a detected image can be decrypted \cite{Feber2013,Feber20142,Rotenberg2014,grosjean2010,caselli2015}. 
 This \textit{a priori} knowledge concerns either information on the structural symmetries, the collection properties of the probe, or knowledge about the investigated field components. The distinction without \textit{a priori} knowledge is much more challenging. Nonetheless, the necessity to have such image reconstruction ability available was mentioned just recently in a review as an important open issue \cite{Rotenberg2014}. To solve this issue, we provide here a methodology to resolve the coupling and therewith provide the ability to reconstruct the entire vectorial electro-magnetic field information from polarization and phase resolved measurements without any prior knowledge. 
 
 Our results reside on the theoretical discussion of the image formation process in aperture based scanning near field optical microscopy. In particular, we discuss the coupling process of the field components to be detected with the eigenmodes sustained by the aperture of the SNOM-tip. The SNOM-tip is treated as a metal coated fiber. Hence, this approach is in close analogy to coupling problems studied in the context of classical fiber optics and provides a slightly alternative description when compared to the pioneering works presented in the past \cite{gre1,Porto2000,VanLabeke1993,Hecht2000,walford2002}. Our approach can be viewed as a special case of the general description provided within the early theories. The advantage of our treatment relies in the simple interpretation of measured quantities. This permits to identify the eigenmodes supported by the aperture as the coherent vectorial point-spread-functions of the aperture SNOM. Analysing these eigenmodes provides insights into the image formation process. We stress that the properties of the eigenmodes, depending on the geometry and the material composition of the tip, will have a strong impact on the image formation process. In particular, probes can be engineered to be sensitive to either the transverse electric or the magnetic components of the investigated field. 
 
 The paper is divided into two parts. In Section \ref{S1} the image formation and especially the conditions leading to the dominant detection of either the electric or the magnetic field components will be discussed. In Sec.~\ref{S3} we focus on polarization and phase resolved measurements obtained by a heterodyne detection scheme and discuss the methodology to solve the inverse imaging problem. We provide an algorithmic approach to retrieve the four in-plane electro-magnetic field components from the two polarization- and phase-resolved measured signals. \\
  \section{Properties of the image formation process} \label{S1}
       \begin{figure}[h]
       \includegraphics[width=1\linewidth]{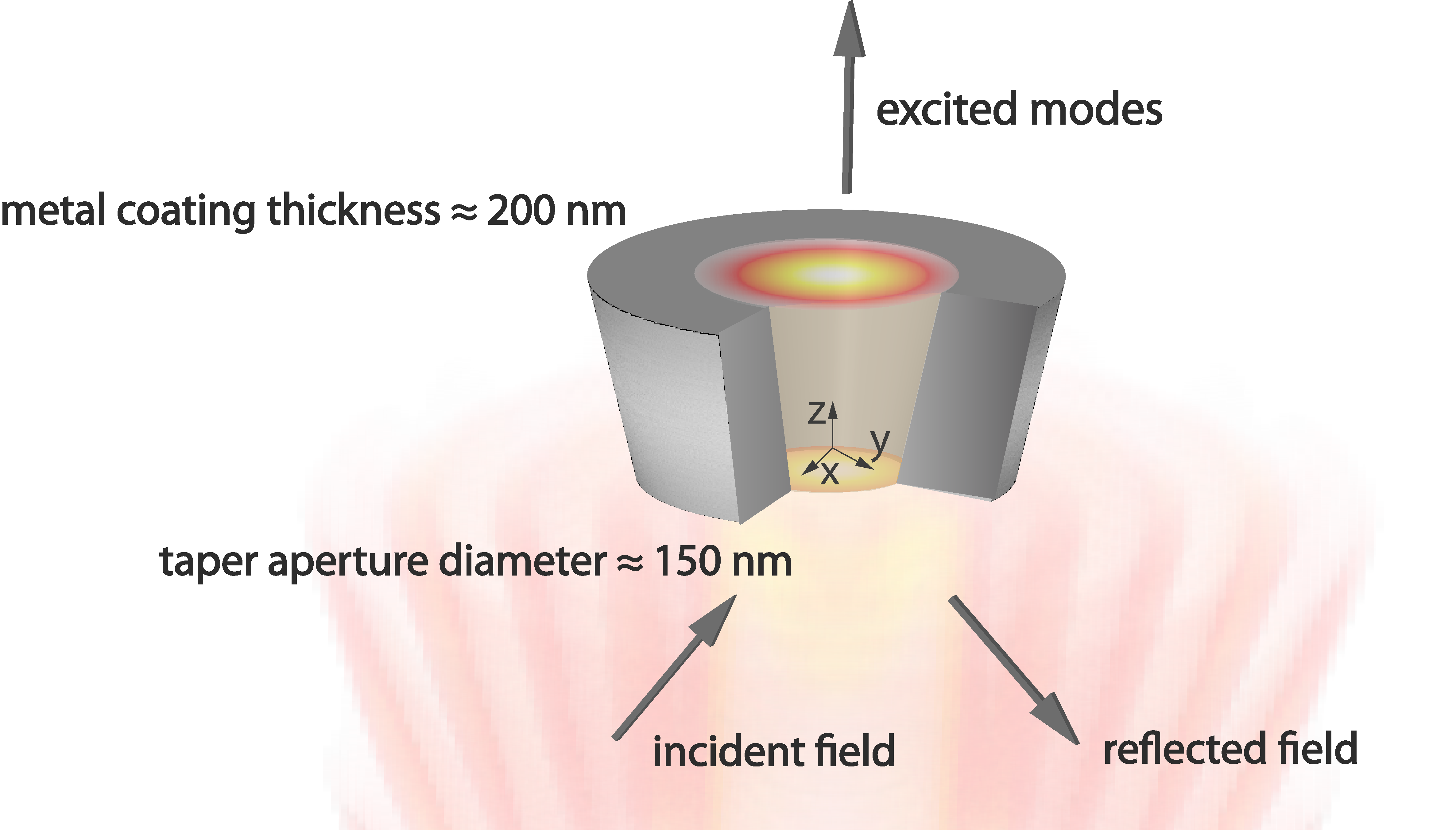}%
       \caption{Sketch of a typical probe used for aperture SNOM measurements in the visible wavelength range. The investigated field incident on the probe excites the modes supported by the aperture. The aperture should be understood as a metal coated fiber with the dimensions of the apex of the tip. The complete coupling process of the externally investigated field into the probe takes place only via the aperture, since the $ 200 $ nm thick metal coating prohibits any side coupling of light into the probe.
        \label{SNOM}}
      \end{figure}
 Understanding the coupling process of the investigated field, which carry structural information about the sample, into propagative modes in the fiber taper is at the heart of the image formation in an aperture SNOM. The corresponding probe consists of a tapered dielectric optical fiber surrounded by a metal coating. In order to avoid any side coupling of light into the probe, the thickness of the metal coating is much larger than the skin depth at the wavelengths of interest. A subwavelength  aperture is milled at the end of this taper (see Fig. \ref{SNOM}). Due to the thick metal coating the coupling process of the investigated field into the tapered fiber occurs only via the aperture at the end of the taper. This suggests that the aperture itself can be understood as the converter for the near into far fields that are guided by a fiber to the detector. To model the coupling process  between the investigated field and the aperture at the apex of the SNOM tip, we treat the aperture as a cylindrical waveguide and calculate its  eigenmodes. The excitation process of these modes by the external field to be investigated, reveals all crucial information about the detected images. The derivation of the mathematical description of this excitation process is presented  within Appendix \ref{model} and is  in close analogy to the early and pioneering works \cite{gre1,Porto2000,VanLabeke1993,Hecht2000,walford2002}. 
  
\begin{figure}[h]
     \includegraphics[width=1\linewidth]{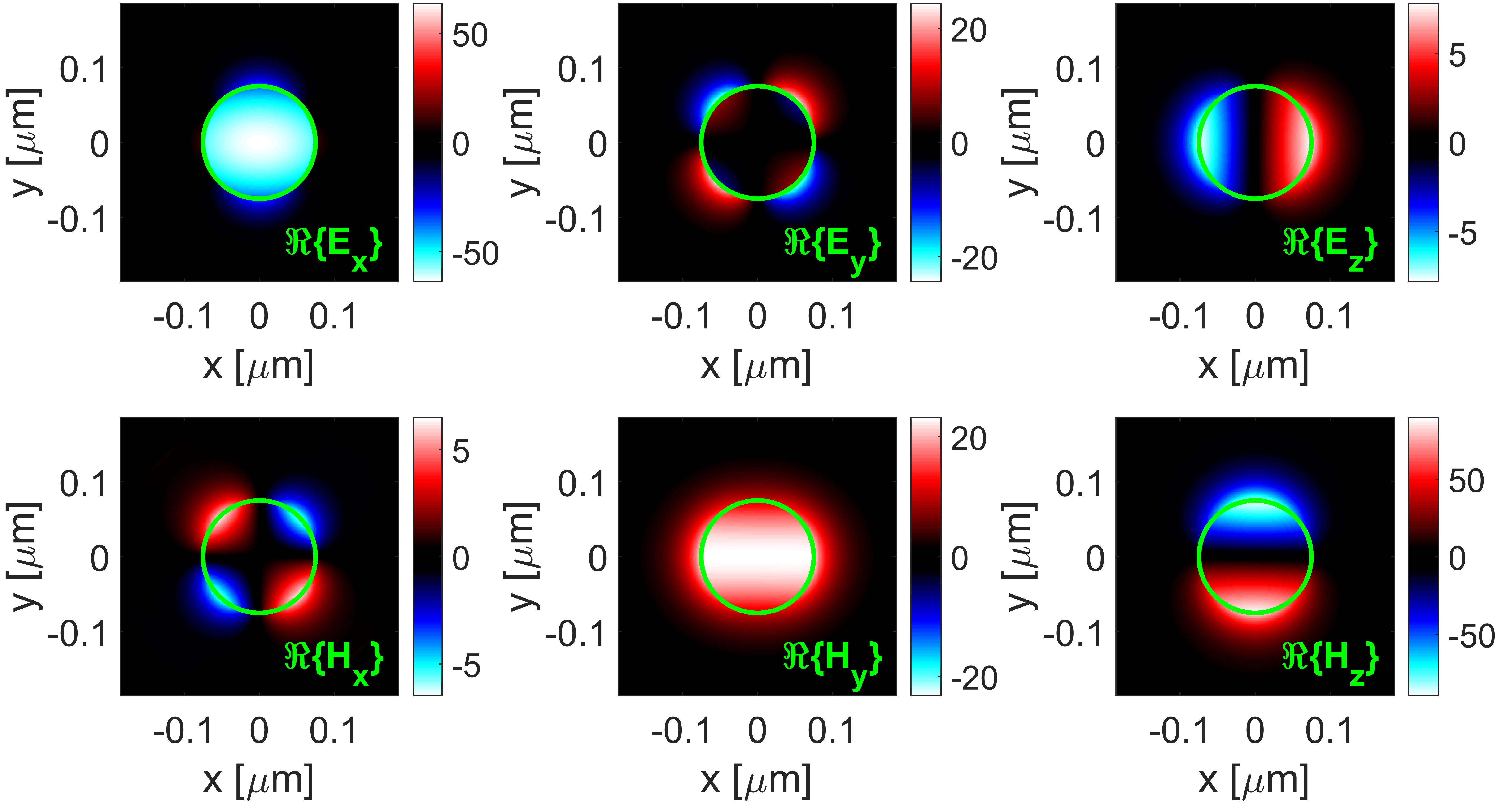}%
     \caption{Real part of the mainly $x$-polarized mode accessible in an aperture at a wavelength of 663 nm. The magnetic field components were multiplied by the free-space impedance $ |Z_0| $, for better comparability of the field magnitudes. The aperture is made of fused-silica glass with a refractive index of $ n=1.5 $ surrounded by a $ 200 $ nm thick gold coating. The $ e_x $ and $ h_y $ fields have a mirror-symmetry with respect to both coordinate axes $ x $ and $ y $, whereas the $ e_y $ and $ h_x $ fields have an odd symmetry. The second accessible mode in the aperture is the mainly $y$-polarized one, in which the symmetries are reversed compared to the mainly $x$-polarized mode.
      \label{Mode}}
    \end{figure}

 Due to the deep subwavelength size of the aperture, it supports only two propagative modes. With \textit{z} as the principal propagation direction, these are the mainly $x$- and $y$-polarized states of the generalized $ HE_{11} $ - modes in the aperture (see Fig.~\ref{Mode}) \cite{novotny1994,novotny}.  The excitation coefficients $ t_x $, $ t_y $ according to the forward propagating $ (+) $ fundamental $ x- $ and $ y- $ polarized modes $ \textbf{e}^{+}_{1,2}(x,y) $ and $ \textbf{h}^{+}_{1,2}(x,y) $ stimulated by an external field can be calculated (see Appendix \ref{model} for details) as 
   \begin{eqnarray}
     t_x(x,y)&=&\int_{\mathcal{R}^2} [\textbf{e}_{1}^{-}\small(\tilde{x}-x,\tilde{y}-y)\times\textbf{H}(\tilde{x},\tilde{y})-\textbf{E}(\tilde{x},\tilde{y})\times\textbf{h}_{1}^{-}(\tilde{x}-x,\tilde{y}-y)]\,\textbf{n}_z\,d\tilde{x}\,d\tilde{y}\, , \label{inv21}\\
   t_y(x,y)&=&\int_{\mathcal{R}^2} [\textbf{e}_{2}^{-}\small(\tilde{x}-x,\tilde{y}-y)\times\textbf{H}(\tilde{x},\tilde{y})-\textbf{E}(\tilde{x},\tilde{y})\times\textbf{h}_{2}^{-}(\tilde{x}-x,\tilde{y}-y)]\,\textbf{n}_z\,d\tilde{x}\,d\tilde{y}\ .\nonumber
        \end{eqnarray}
 Here, $ \textbf{E}(\tilde{x},\tilde{y}) $ and $ \textbf{H}(\tilde{x},\tilde{y}) $ correspond to the time harmonic electro-magnetic field components of the investigated field  in the plane $ z=0 $ defined by the height of the aperture and $ (x,y)^T $ defines the position of the tip (see Fig. \ref{SNOM}). The time resolved harmonic fields depending on the frequency $ \omega $ are given by $ \mathcal{E}(x,y,z=0,t)= \textbf{E}(x,y)\,e^{-i\,\omega\,t}$ and $ \mathcal{H}(x,y,z=0,t)= \textbf{H}(x,y)\,e^{-i\,\omega\,t}$.   Consequently a detected image is identified via the guided power of the modes supported by the aperture \cite{sny}
       \begin{eqnarray}
                \mathcal{I}(x,y)=&&|t_x(x,y)|^2+|t_y(x,y)|^2 \nonumber\\
                =&&\left |\int_{\mathcal{R}^2} [\textbf{e}_{1}^{-}\small(\tilde{x}-x,\tilde{y}-y)\times\textbf{H}(\tilde{x},\tilde{y})-\textbf{E}(\tilde{x},\tilde{y})\times\textbf{h}_{1}^{-}(\tilde{x}-x,\tilde{y}-y)]\,\textbf{n}_z\,d\tilde{x}\,d\tilde{y}\right |^2+ \label{bildges}\\
                    &&\left |\int_{\mathcal{R}^2} [\textbf{e}_{2}^{-}\small(\tilde{x}-x,\tilde{y}-y)\times\textbf{H}(\tilde{x},\tilde{y})-\textbf{E}(\tilde{x},\tilde{y})\times\textbf{h}_{2}^{-}(\tilde{x}-x,\tilde{y}-y)]\,\textbf{n}_z\,d\tilde{x}\,d\tilde{y} \right |^2 \nonumber.
                           \end{eqnarray}
 This result is in analogy to the more general scenario presented in Refs.~\cite{Porto2000,walford2002}. Expressions from these pioneering work are recovered if the eigenmodal-fields considered here are substituted with the reciprocal fields obtained by evaluating two scenarios fulfilling the requirements of reciprocity. The theory developed in Refs.~ \cite{Porto2000,walford2002} were successfully applied to explain SNOM measurements \cite{bakker2015,Feber2013,Feber20142,sinev2015}. The advantage of our formulation above is the interpretation of the eigenmodes sustained by the aperture as the coherent point-spread-function of the aperture SNOM. Then, the result can be regarded in terms of classical microscopy under coherent illumination, where the image is formed correspondingly as the convolution between the coherent point-spread function and the field of the investigated structure of interest in the case of an isoplanatic condition \cite{gross2005handbook,singer2006handbook}. However, in contrast to classical microscopy, the point spread function in aperture based SNOM is vectorial, encoding not only electric but also magnetic field components in a detected image. From the expression above it becomes evident that the dominant influence of either the electric or the magnetic field components in a detected image not only depends on the investigated field but also on the electro-magnetic properties of the two modes accessible in the aperture. 
 
 The further distinction of individual field components contributing to a detected image in Eqn.~\ref{bildges} is impossible, in general, due to the encryption through the convolution process. To get a deeper insight into the relation between investigated field components and a detected image, we treat the problem from now on in the quasi-static limit. In the quasi-static limit the investigated field is assumed to be constant across the aperture. The advantage of this approximation is the ability to derive semi-analytical expressions for the measured intensity. They enable the identification of the measured field components in a detected image. This treatment departs the discussion of the image formation process in aperture SNOMs from its super-resolving character. Nevertheless, these assumptions have a considerable practical impact, as aperture SNOM's are often used to study non-propagating field components, i.e. related to plasmonic structures, where super-resolution is not necessarily required. Often in these cases fundamental insights were gathered while comparing measured  images  with simulated field-distributions of the investigated structure \cite{dev1,dev2,Angelaneu,ang1,bur2,ang2,bou1,bur1,bur4,uebel2012,Feber2013,Denkova2013,Denkova2014,koh1,kihm1,Kihm2,bur2,Feber20142,Rotenberg2014}. The ability to directly relate measured images with simulated field components led to exceptional results,  but is, in a strict sense, only possible within the validity of the quasi-static limit.
 
 While assuming the investigated field components to be constant across the aperture ($\textbf{E}(\tilde{x},\tilde{y})=\textbf{E}$ and $ \textbf{H}(\tilde{x},\tilde{y})=\textbf{H}$), Eqn. \ref{bildges} can be rewritten. The excitation strength corresponding to the mainly x-polarized mode within the aperture is given by 
    \begin{eqnarray}
    t_x(x,y)&=&\int[\textbf{e}_{1}^{-}\times\textbf{H}-\textbf{E}\times\textbf{h}_{1}^{-}]\,\textbf{n}_z\,d\tilde{x}\,d\tilde{y}\nonumber\ ,\\
    &=& H_y\int e^{-}_{1_{x}}d\tilde{x}d\tilde{y}-H_x\underbrace{\int e^{-}_{1_{y}}d\tilde{x}d\tilde{y}}_{=0} 
    -E_x\int h^{-}_{1_{y}}d\tilde{x}d\tilde{y}+E_y\underbrace{\int h^{-}_{1_{x}}d\tilde{x}d\tilde{y} }_{=\ 0}\label{quasieqn2}\ ,\\
    &=& H_y\underbrace{\int e^{-}_{1_{x}}\,d\tilde{x}\,d\tilde{y}}_{\overline{e_{1_{x}}}}-E_x\underbrace{\int h^{-}_{1_{y}}\,d\tilde{x}\,d\tilde{y}}_{\overline{h_{1_{y}}}} \nonumber\ , \\
    &=& H_y\,\overline{e_{1_{x}}}\,-\,E_x\,\overline{h_{1_{y}}} \label{quasieqn1}\ .
    \end{eqnarray} 
    In Eqn.~\ref{quasieqn2} two integrals vanish due to the odd symmetry of the corresponding mode-components (see Fig.~\ref{Mode}). In a next step, the value $ \overline{h_{1_{y}}} $ of the magnetic mode component of the mode in the aperture is expressed by the value $ \overline{e_{1_{x}}} $ of the electric mode component of the aperture
    \begin{eqnarray}
    \overline{h_{1_{y}}}&=&Z_M\,^{-1}\,\overline{e_{1_{x}}}\label{qe1}\ ,\\
    Z_M&=&\frac{\overline{e_{1_{x}}}}{\overline{h_{1_{y}}}} \nonumber\ .
    \end{eqnarray}  
    At this point, the mode impedance $ Z_M $ has been introduced. It expresses the ratio between the electric and magnetic mode components. By repeating the calculations for the mainly y-polarized mode accessible in the aperture, assuming a cylindrical symmetry of the probe, a final expression for the measured intensity can be derived that reads as
  \begin{eqnarray}
    \mathcal{I}&=&\vert t_x\vert^2+\vert t_y\vert^2 \ ,\ \ \ \ \ \ \ \,   \nonumber \\
    &=&|\textbf{H}_{\perp}|^2+|Z_M|^{-2}|\textbf{E}_{\perp}|^2- 2|Z_M|^{-1}|E_x||H_y|\cos(\phi_{Hy}-\phi_{Ex}-\phi_{Z_M})\nonumber\\
    &&\ \ \ \ \ \ \ \ \ \ \ \ \ \ \ \ \ \ \ \ \ \ \ \ \ \ \  \ \ \ \ \ \ \ \, \,-2|Z_M|^{-1}|H_x||E_y|\cos(\phi_{Hx}-\phi_{Ey}-\phi_{Z_M}) \label{image3}\ .
  \end{eqnarray} 
 Here $ \phi $ denotes the phase values of the complex quantities. Proportionality constants were neglected, since they have no qualitative influence on the image formation.
  \begin{figure}[h]
      \includegraphics[width=1\linewidth]{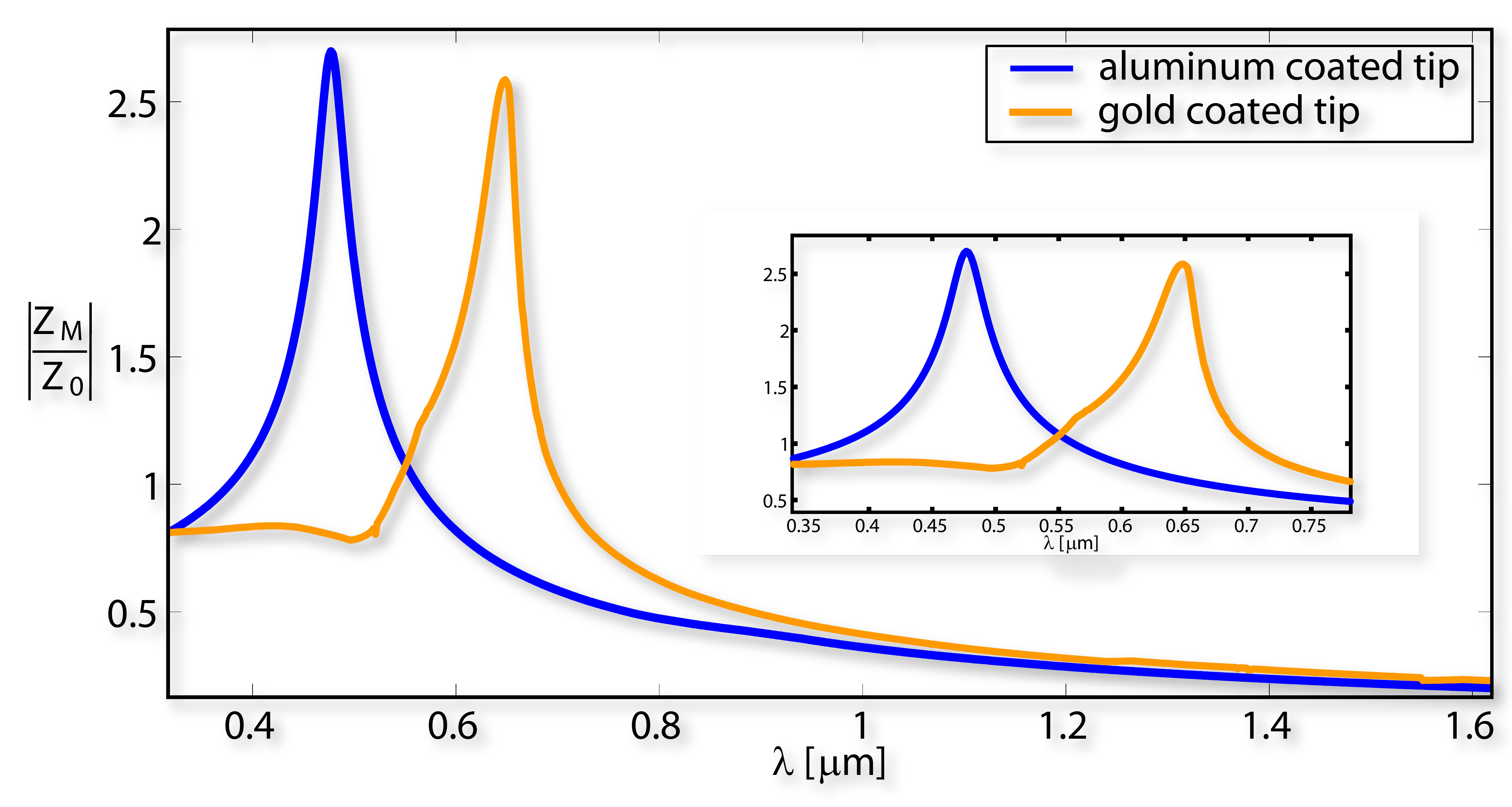}%
      \caption{Spectrally resolved mode-impedance factor $ \left |\frac{Z_M(\lambda)}{Z_0}\right | $ for 2 commercially available probes normalized by the free-space impedance $ Z_0=376\, \Omega$. Both tips share the same geometrical parameters, i.e. the aperture diameter is $ D=150 $ nm, the core is made from fused silica with $ n=1.5 $, and the core is covered with a $ 200 $ nm thick metal coating. The tips differ only by their coating-material. Both probes show resonances of the mode-impedance in the visible wavelength range, which are slightly shifted against each other. The resonance of the gold coated tip lies in the range of commonly used wavelength for near field measurements with such aperture diameters. In this regime, around $ \lambda=650 $ nm, gold coated probes tend to be much more sensitive to magnetic field components in a detected image when compared to aluminum coated tips.}
       \label{imp}
     \end{figure}
 
 The result obtained in Eqn.~\ref{image3}  is one key finding of this work. The expression is applicable to directly explain a possible measurement and provides insights into the properties of the image formation process in aperture SNOMs. Specifically, two things become evident. On the one hand longitudinal field components, namely $ |E_z|$ and $|H_z|$, cannot be detected. This is in agreement with past results \cite{dev1,dev2,kihm1,Kihm2,Denkova2013,Denkova2014,ang1,bur2,ang2,bou1,bur1,bur4}. Hence, the aperture SNOM is a complementary technique to the cross-polarized scattering-type SNOM, where dominantly $|E_z|$ is measured \cite{Esslinger2012,Kihm2,ez1,ez2}. On the other hand, the mode impedance $ |Z_M| $ influences the sensitivity of the aperture SNOM to detect either electric or magnetic field components. Therefore, it describes the impact of the probe itself on the measurement. This will be discussed in more detail in the following.
 
 Equation \ref{image3} has two individual contributions which are called the amplitude and the phase term in the following. Since these parts influence a detected image differently, their properties will be discussed individually. In a first step of discussion, the phase term will be neglected. 
 
 The amplitude term, $ |\textbf{H}_{\perp}|^2+|Z_M|^{-2}|\textbf{E}_{\perp}|^2 $, generally expresses the possibility to measure simultaneously the electric and magnetic field components. Their relative contribution is dictated by the impedance of the mode sustained by the aperture of the tip. 
 To analyse the relative ratio of electric and magnetic field components contributing to a detected image, the ratio $ \Gamma=\frac{\left |{ { Z_M}}\right  |^{-2} \left |{ \textbf{\scriptsize   E}}_{\perp}\right |^2}{\left |\textbf{\scriptsize   H}_{\perp}\right |^2} $ is evaluated.  By replacing $\frac{ \left  |\textbf{\scriptsize E}_{\perp}\right |}{\left |\textbf{\scriptsize H}_{\perp}\right |} $ with the transverse field impedance $ |Z_{\perp} |$, the following result is obtained \begin{eqnarray}
  \Gamma=\frac{\left |Z_\perp\right |^2}{\left |Z_M\right  |^2}\ \ \ . \label{ratio}
 \end{eqnarray}  This is the ratio between the investigated field- and the mode impedance that determines the relative influence of the individual electric and magnetic field components in a detected image.
 
 Since the mode impedance $\left  |Z_M\right  | $  depends on the geometry and the material of the tip, the detection properties of an aperture SNOM are strongly influenced by the choice of the probe. This is exemplified in Fig.~\ref{imp}, where the calculated spectrally resolved mode impedance $ Z_M(\lambda) $ is shown for commercially available gold and aluminum coated probes with an aperture diameter $ D $ of $ D=150 $ nm. The numerical analysis of the eigemodes guided in those apertures has been done with COMSOL. It can be seen that both tips show strong resonances in the visible, leading to spectral variations in their detection characteristics. While the resonance for the gold coated probe lies at a wavelength of approximately $ \lambda\approx650 $ nm, the resonance for the aluminum coated probe is shifted and lies at $ \lambda\approx470 $ nm. These resonances originate from the excitation of localized surface plasmon polaritons in the aperture. This renders the tips not just sensitive on the aperture diameter but also on the material. Tailoring these resonances allows to design probes that are dominantly sensitive to either the electric or magnetic field components of light in a specific spectral region. By using Babinet's principle, the resonances of those apertures can be linked in lowest order approximation to resonances of cylindrical metallic nanoparticles, for which analytical means exist to predict the spectral position of the resonance \cite{Rockstuhl2008,Abajo2007}.  
 
 By assuming an electro-magnetic field with the impedance of free space $ Z_0=376\, \Omega $ to be investigated at a wavelength of $ \lambda\approx650 $ nm, the image  detected with the proposed gold  coated tip would then strongly be influenced by magnetic field components. It would be the opposite for the aluminum coated tip, where a detected image would be dominated by electric field components.
 
 Generally, for arbitrary electro-magnetic fields the investigated field impedance $ |Z_{\perp}(x,y)| $  is a spatially varying quantity that depends on the local optical quantities of the measured field. In a measurement, this behaviour causes a local variation of the dominant influence of either the electric or the magnetic field components in a detected image. These behaviours already have been investigated experimentally, where the relative information content in a measurement related to electric and magnetic field parts was examined \cite{Feber2013,Feber20142,Rotenberg2014}. There it was shown that both the explicit choice of a probe and the relative position on the sample changes the detection characteristics.  Here, we explain this experimental finding with Eqn.~\ref{ratio} as the local variation of the  impedances of the investigated field and the modes accessible in the aperture.
 
 The ability to investigate electro-magnetic fields with spatially distinguishable electric and magnetic field components is challenging, while strictly fulfilling the requirements of the quasi-static limit. It requires the investigation of electro-magnetic fields for which the intensity related to the time-averaged Poynting-Vector $ <\textbf{S}>=\frac{1}{2}\Re\left \{\textbf{E}\times\textbf{H}^*\right \} $ is not proportional to the local electric energy density  $ |\textbf{E}|^2\not \propto <\textbf{S}> $. Possibilities to investigate electro-magnetic fields with a strongly distinguishable character of the individual electric- and magnetic components could  be done by evaluating strongly focused radially polarized fields or alternatively standing wave patterns \cite{dorn2003,novotny2001}.  
 
 Additionally, an unexpected influence on a measurement can be caused  by the phase term $ 2|Z_M|^{-1}|E_x||H_y|\cos(\phi_{Hy}-\phi_{Ex}-\phi_{Z_M})
  +2|Z_M|^{-1}|H_x||E_y|\cos(\phi_{Hx}-\phi_{Ey}-\phi_{Z_M}) $ contributing within the image formation process. Here, phase differences between the local electric and magnetic field components influence a measurement. This effect can lead to unexpected influences in a detected image, making them difficult to interpret. However, for the investigated fields $ \textbf{E}, \textbf{H} $ which additionally obey a slowly varying envelope approximation \textit{SVEA},  the phase differences $ \left (\phi_{Hy}-\phi_{Ex}-\phi_{Z_M}\right )$ and  $ \left (\phi_{Hx}-\phi_{Ey}-\phi_{Z_M} \right )$ remain approximately constant across the scanning area.   The phase term then affects a detected image only by a decrease in contrast. Although this seems to be a restriction, it actually constitutes to be valid in many practical situations, as, to the best of our knowledge, emerging influences have not been reported in literature.  
 
 \section{Inverse imaging problem} \label{S3}
 In the previous section we discussed the image formation process in aperture SNOMs by assuming non-interferometric measurements, i.e. only the intensity of the modes have been considered. To go beyond, we present in the following an algorithm to reconstruct the complete vectorial electro-magnetic field from a phase resolved measurement of the two excited eigenmodes. Such a measurement technique was developed during the last  years and enables the measurement of a polarization-resolved signal in both amplitude and phase by using a heterodyne detection scheme \cite{het2}. The establishment of a heterodyne detection scheme for aperture based scanning near field measurements led to outstanding results \cite{Feber2013,Feber20142,bur1,bur2,bur3,bur4,Rotenberg2014}. 
  \subsection*{Polarization and phase resolved signal}
 It is natural to assume that the polarization- and phase-resolved measured signals correspond to the excitation strengths of either the mainly $x$- or $y$-polarized mode $ t_{x,y}(x,y) $ accessible in the aperture according to Eqn. \ref{inv21}. Within the quasi-static-limit the recorded signals correspond then either to  a superposition of $ E_x,H_y $ or $ E_y,H_x $ (see Eqn.~\ref{quasieqn1}), being in agreement with results presented in Ref.~ \cite{Feber2013,Feber20142}. Correspondingly, within a measurement the four transverse electro-magnetic field components are obtained within the two polarization resolved signals. The discrimination of the individual field components in a detected image is challenging and requires a second set of equations or assumptions to resolve the discrepancy of an under-determined system. This has already been shown by exploiting either an underlying symmetry, collection properties of the probe, or {\it a priori} knowledge of the investigated field components \cite{Feber2013,Feber20142,Rotenberg2014,grosjean2010}. Here, we would like to discuss this issue in a more general way, by solving the inverse imaging problem without prior knowledge about the structure or related symmetries. In the following, the algorithmic approach to unravel the distinct field components will be presented.
 
 In the following we only have one assumption. Once an eigenmode is excited at the apex, the amplitude  of the mode  is not influenced by any coupling process while it propagates through the fiber to a detector. Due to imperfections in an experimental setup, e.g. by fiber-bends, this might not be guaranteed. A coupling between the orthogonal modes due to imperfections might occur, causing a mixing of the individual detection paths. However, in reality the problem can be experimentally solved by a prior calibration of the setup while investigating the transmission of linearly polarized light. By independently inspecting linearly $x$- and $y$-polarized fields, the Jones-Matrix of the SNOM can be determined. From the knowledge of these matrix-entries in amplitude and phase, the polarization distortion introduced by weak disturbances within the measurement system can be analysed. It is then possible to remove the distortions numerically to infer on the actual quantities of interest. Eventually, such calibration allows direct access to the complex, polarization resolved excitation coefficients according to Eqn.~\ref{inv21}. With this assumption we can continue to develop our field reconstruction algorithm. 
        \subsection*{Full vectorial field reconstruction}
 In the following, we present an algorithm to reconstruct the investigated field components from a phase resolved measurement.  By introducing the Fourier-transformation 
      	\[   \tilde{h}\left (\textbf{k}_\perp\right )=\mathcal{F}\left \{h\left (\textbf{r}_\perp\right ) \right \}=\frac{1}{2\pi}\int_{\mathcal{R}^2}h\left (\textbf{r}_\perp\right )e^{-i\left (k_xx+k_yy\right )} d\textbf{r}_\perp \]      and its inverse  \[ h\left (\textbf{r}_\perp\right )=\mathcal{F}^{-1}\left \{ \tilde{h}\left (\textbf{k}_\perp\right ) \right \}=\frac{1}{2\pi}\int_{\mathcal{R}^2}\tilde{h}\left (\textbf{k}_\perp \right )e^{i\left (k_xx+k_yy\right )} d\textbf{k}_\perp,\]  Eqn.\ref{inv21} can be written in Fourier domain as
      \begin{small}
      	    \begin{eqnarray}
      	        \tilde{t}_{x}\left (\textbf{k}_\perp \right )=\,& \left [\tilde{\textbf{e}}_{1}^{-}\left (\textbf{k}_\perp \right )\,\times\,\tilde{\textbf{H}}\left (\textbf{k}_\perp \right )-\tilde{\textbf{E}}\left (\textbf{k}_\perp \right )\,\times\,\tilde{\textbf{h}}_{1}^{-}\left (\textbf{k}_\perp \right )\right ]\textbf{n}_z\, ,  \label{inv2}\\
   \tilde{t}_{y}\left (\textbf{k}_\perp \right )=\,&\left [\tilde{\textbf{e}}_{2}^{-}\left (\textbf{k}_\perp \right )\,\times\,\tilde{\textbf{H}}\left (\textbf{k}_\perp \right )-\tilde{\textbf{E}}\left (\textbf{k}_\perp \right )\,\times\,\tilde{\textbf{h}}_{2}^{-}\left (\textbf{k}_\perp \right )\right ]\textbf{n}_z\,.     
 \notag
      \end{eqnarray}
           \end{small}
   Here quantities $ \tilde{j}(\textbf{k}_{\perp}) $ are the Fourier-representations of the respective functions $ j(\textbf{r}_{\perp}) $ in real space. Equation \ref{inv2} can be understood as an under-determined system of equations, where it is intended to infer from the knowledge of the measured and Fourier transformed mode coefficients $ \tilde{t}_{x,y} $  onto the 4 unknown investigated field components $ \tilde{E}_x,\tilde{E}_y,\tilde{H}_x,\tilde{H}_y $ in Fourier domain. By exploiting free space Maxwell's Equations in the angular frequency domain, this limitation can be overcome and the apparent magnetic field components $ \tilde{H}_x,\ \tilde{H}_y $  can be equivalently expressed by the electric field components $ \tilde{E}_x,\ \tilde{E}_y $. This results in a well determined system of equations
  \begin{eqnarray}
  \tilde{t}_x&=&\underbrace{(\gamma_3\tilde{e}^{-}_{1_{x}}-\gamma_1\tilde{e}^{-}_{1_{y}}-\tilde{h}^{-}_{1_{y}})}_{M_{11}}\tilde{E}_x+\underbrace{(\gamma_4\tilde{e}^{-}_{1_{x}}-\gamma_{2}\tilde{e}^{-}_{1_{y}}+\tilde{h}^{-}_{1_{x}})}_{M_{12}}\tilde{E}_y\nonumber,\\
  \tilde{t}_y&=&\underbrace{(\gamma_3\tilde{e}^{-}_{2_{x}}-\gamma_1\tilde{e}^{-}_{2_{y}}-\tilde{h}^{-}_{2_{y}})}_{M_{21}}\tilde{E}_x+\underbrace{(
  \gamma_4\tilde{e}^{-}_{2_{x}}-\gamma_{2}\tilde{e}^{-}_{2_{y}}+\tilde{h}^{-}_{2_{x}})}_{M_{22}}\tilde{E}_y,\label{inv3}\\
  \gamma_1&=&-\frac{1}{k_0Z_0}\frac{k_yk_x}{k_z},\ \ \ \ \ \ \ \ \ \ \ \ \ \ \ \gamma_{2}=-\frac{1}{k_0Z_0}\left(\frac{k_y^2}{k_z}+k_z\right)\nonumber,\\
  \gamma_3&=&\frac{1}{k_0Z_0}\left(\frac{k_x^2}{k_z}+k_z\right),\ \ \ \ \ \ \ \ \ \ \gamma_4=\frac{1}{k_0Z_0}\frac{k_yk_x}{k_z}\nonumber,\\
  k_z&=&\sqrt{\left (\frac{2\pi}{\lambda}\right )^2-k_x^2-k_y^2}\ .\nonumber
   \end{eqnarray}
   The individual matrices $M_{ik}\left (\textbf{k}_\perp \right )$ can be interpreted as the vectorial transfer-function of the SNOM-tip. It determines not just the spatial resolution but also the achievable contrast. These quantities are determined by the two fundamental modes accessible in the aperture.
   
 The system of Eqns. \ref{inv3} can be inverted independently for every transverse wave vector component $\textbf{k}_\perp$ resulting in the two independent field components $\tilde{E}_x\left (\textbf{k}_\perp \right )$  and $\tilde{E}_y\left (\textbf{k}_\perp \right )$ in Fourier-representation. From these information the remaining four field components $\tilde{E}_z\left (\textbf{k}_\perp \right )$, $\tilde{H}_x\left (\textbf{k}_\perp \right )$, $\tilde{H}_y\left (\textbf{k}_\perp \right )$ and $\tilde{H}_z\left (\textbf{k}_\perp \right )$ can be derived using Maxwell's equations for the homogeneous free space. By inversely Fourier transforming the independent field components the investigated field can be reconstructed. The achievable resolution in the reconstructed investigated field components is limited by the accessible spatial frequencies in the different $M_{ik}\left (\textbf{k}_\perp \right )$.
 
 This algorithm has the capability to fully reconstruct all vectorial components of the investigated field. This approach allows to extract information obtained by an aperture SNOM measurement without \textit{a-priori} knowledge and can provide the ability to give novel insights in nano-optical research activities due to the ability to infer onto the complete vectorial information from a measurement. 
 
 From the numerical point of view, the only problem concerns the different discretizations of the investigated field and the modes supported by the aperture. Due to the nanoscopic dimensions of the aperture and the macroscopic extension of the investigated fields, the numerical sampling is different and has to be unified to evaluate Eqn.~\ref{inv3}. Due to the usual extreme ratio of the two different scales, approaches like zero padding can cause difficult memory issues and prevent the calculation of the investigated field components. To prevent the problem, chirp z-transformations can be used to unify the grids without causing memory problems \cite{singer2006handbook,gro12,gro2,gro3}.       
   \section{Conclusion}
 In conclusion, we discussed the properties of the image formation and the inverse imaging problem in aperture-based scanning near field optical microscopy. In particular, we assessed the properties of aperture SNOM probes that lead to a dominant sensitivity to either the electric or the magnetic field components in a detected image. We proved that these detection characteristics are strongly linked to the properties of the modes sustained by the aperture of the SNOM. These properties can be tailored by changing the geometry and material composition of the probe. It constitutes the means to design probes that can detect specific field components on demand. In particular, we proposed a specific probe that is either coated by aluminum or gold that measures dominantly either the electric or magnetic field components at a specific wavelength. In addition, we provided a methodology to solve the inverse imaging problem to extract the complete vectorial field information from phase resolved measurements. 
 \section*{Acknowledgment}
 S.S. would like to thank Norik Janunts, Bayarjargal Narantsatsralt for helpful discussions and experimental measurements.
 Support by the German Federal Ministry of Education and Research (KoSimO, PhoNa) and by the Thuringian State Government (MeMa) is acknowledged. We gratefully acknowledge partial financial support by the Deutsche
 Forschungsgemeinschaft (DFG) through CRC 1173 and Open Access Publishing Fund of Karlsruhe Institute of Technology.
 \appendix
 \section{Model} \label{model}
 Here, we discuss in detail the coupling process of the investigated field components with the modes supported by the aperture of the probe. 
 
 Due to the deep subwavelength size of the aperture, it supports only two propagative modes. These are the mainly $x$- and mainly $y$-polarized states of the generalized $ HE_{11} $ - modes in the aperture (see Fig.~\ref{Mode})\cite{novotny1994,novotny}. To conclude from the knowledge of these excitation strengths of the aperture modes on an image that can be measured by a detector, not only the aperture but also the complete measurement setup needs to be considered at the same time. We require here that the excitation strengths of the aperture modes are not modified by any coupling process with any other mode at a later stage of propagation through the system. This requires adiabatic modifications to the geometry of the fiber, e.g. while being tapered or being bended. Then, the modal amplitudes at the position of the detector are linearly related to the ones in the aperture. 
 
 The total power guided by the fiber to the detector is proportional to the absolute square of the excitation strengths of the supported modes\cite{sny}. Generally, the power can be written  as $ P=\sum_{i=1}^{\infty} |t_{i_\mathrm{\;detector}}|^2$, assuming orthogonal and power-normalized modes. In the considered case solely two modes are excited which are linearly related to the ones in the aperture. It follows that the guided power at the position of the detector can be written as $ P\propto |t_{x_\mathrm{\;aperture}}|^2+|t_{y_\mathrm{\;aperture}}|^2$. Consequently, we identify a detected image $ \mathcal{I} $ via the guided power at the position of the detector from the knowledge of the excitation strengths in the aperture  as $ \mathcal{I}=|t_{x_\mathrm{\;aperture}}|^2+|t_{y_\mathrm{\;aperture}}|^2 $.
 
 These excitation strengths depend on the position of the probe relative to the sample. Consequently, a detected image, when the probe scans the sample at constant height, should be written as \begin{eqnarray}
   \mathcal{I}(x,y)=|t_{x}(x,y)|^2 +|t_{y}(x,y)|^2 \ .\label{bild0}
   \end{eqnarray} Here the excitation strengths of the mainly linearly polarized modes in the aperture were renamed as $ t_{x,y}:=t_{x,y_\mathrm{\; aperture}} $. These amplitude coefficients correspond to the projection of the investigated field onto the modes supported by the aperture. It follows, that the detected image does not correspond to the investigated field components directly, than rather to the mentioned projection.
 
 Keeping in mind the assumptions that were made, this approach simplifies the complex problem of the image formation process in aperture SNOM to the question of understanding the coupling process of the two fundamental modes accessible in the aperture excited by the externally investigated field. This remaining problem can be regarded in terms of classical fiber optics and is in direct analogy to the treatment of splicing and coupling losses in optical fibers.
 
 The incident field will couple power into the two propagative modes in the aperture, but also into the unbound radiation and the evanescent modes supported by the aperture. Additionally, it also causes a partial reflection of the incident field. To extract the coupling coefficients of the system, Maxwell's interface problem has to be solved at the position of the aperture. However, only the excitation coefficients for the guided modes are important. They will be derived in the following. In this derivation we assume monochromatic fields with a time dependency proportional to $ e^{-i\omega t} $.
 
 By formally introducing Dirac's notation, the transverse electro-magnetic field of the $m^\mathrm{th}$  forward-propagating eigenmode in the aperture is denoted as $ \ket{\textbf{T}^{+}_{m}} $, where $ m=1,2 $ corresponds to the mainly $x$- and mainly $y$-polarized mode of interest and any higher $m$ stands for evanescent and radiation modes. The Dirac notation should be understood in the sense of a 4-component vector denoting the transverse mode profile as  $ \ket{\textbf{M}^{+}_{m}}=\left [e_{x_{m}}(x,y),e_{y_{m}}(x,y),h_{x_{m}}(x,y),h_{y_{m}}(x,y)\right ]^\top$, that needs to be computed by any mode solving technique. Consequently, also the investigated incident and reflected field is regarded as a superposition of eigenmodes in free space. In the following  we choose a plane waves basis, where every plane wave is written in Dirac-notation as  $ \ket{\textbf{P}^{\pm}_{\small  \textbf{k}_{\perp}}} $. There, the $ \pm $ sign either denotes forward $ (+) $ or backward $ (-) $ propagating waves. These are the incident and the reflected fields, respectively.  $ \textbf{k}_{\perp} $ stands for the unique transverse wave vector of the plane wave $ \textbf{k}_{\perp}=\left (k_x,k_y\right )^{\top}$.  
 
 The interface problem, describing the coupling of the investigated field to the modes supported by the aperture, requires continuity of all tangential field components at the aperture $ \textbf{E}^{\mathrm{~freespace}}_\perp(x,y,z=0)\stackrel{!}{=}\textbf{E}^{\mathrm{~aperture}}_\perp(x,y,z=0)$ (analogous for the magnetic fields $ \textbf{H}_{\perp} $). This can be written as follows
   \begin{eqnarray}
 \small  \underbrace{\int_{\mathcal{R}^2}
    i_{\tiny \textbf{k}_{\perp}}\,\ket{\textbf{P}^{+}_{{ \tiny \textbf{k}_{\perp}}}}}_{\text{incident field }} + \underbrace{\int_{\mathcal{R}^2}
    r_{\tiny \textbf{k}_{\perp}}\,\ket{\textbf{P}^{-}_{{\tiny \textbf{k}_{\perp}}}}}_{\text{reflected field}}=\underbrace{\int\!\!\!\!\!\!\!\!\!\!\!\;\sum
    t_m\, \ket{\textbf{T}^{+}_{m}}}_{\text{excited aperture modes}} ,  \label{E1}
   \end{eqnarray}
   where $ \int\!\!\!\!\!\!\!\!\!\;\sum $ symbolizes the discrete sum of the finite number of bound modes together with the integration of the infinite number of radiation modes in the system. 
 
 Equation \ref{E1} is fundamental describing any coupling problem between any two structures and their supported eigenmodes. Beneficially, Eqn. \ref{E1} can be solved by exploiting the unique properties of the modes that are governed by an unconjugated reciprocity \cite{lal,sny}. An inner product can be defined as
   \begin{eqnarray}
   \braket{\psi_m|\psi_n}=\int_{\mathcal{R}^{2}}[\textbf{e}_m\times\textbf{h}_n-\textbf{e}_n\times\textbf{h}_m] \textbf{n}_z\,dx\,dy
   \end{eqnarray}
   in between the supported eigenmodes of the structures, where $ \textbf{n}_z $ is the unity vector in z-direction.  The orthogonality relation for modes of the same eigenmodal system reads as
   \begin{eqnarray*}
   \braket{\psi_m^+|\psi_n^-}=\alpha_m^{~2}\,\delta_{mn}\, ,\ &&\  \braket{\psi_m^+|\psi_n^+}=0\, ,\\
   \braket{\psi_m^-|\psi_n^+}=-\alpha_m^{~2}\,\delta_{mn}\, ,\  &&\  \braket{\psi_m^-|\psi_n^-}=0\, ,
   \end{eqnarray*} 
  with $ \alpha_m $ to be the normalization factor.  On the base of this inner product, Eqn.~\ref{E1} can be solved self-consistently requiring the explicit knowledge about the limited number of bound modes but also on the unlimited number of unbound radiation modes. The inclusion of the infinitely extended radiation modes on a numerically truncated grid is unfeasible. Therefore, further truncations depending upon the system to be investigated are required. 
  
 In the considered case  of image formation in collection-mode aperture SNOM, the neglection of the reflection in Eqn.~\ref{E1} is a reasonable assumption. This approximation can be considered as the first order perturbation theoretical approach to describe the system. Due to the deep subwavelength dimensions of the aperture of the tip the reflective response of the tip is weak and has a wide spectral bandwith in the angular plane wave spectrum, justifying the assumptions made. It corresponds to the assumption that the field to be measured is not perturbed by the tip of the SNOM. This assumption was mentioned in Ref.~ \cite{Rotenberg2014} to be mostly valid in practical situations.
    
 By neglecting the reflection in Eqn. \ref{E1}, the problem reduces to the expansion of the investigated field into the modes supported by the aperture. The explicit representation of the investigated field in the plane wave basis is no longer necessary and thus the incident field $ \textbf{E}(x,y) $, $ \textbf{H}(x,y) $ will be renamed in Dirac-notation as $ \ket{\tiny{\begin{array}{c}
          \stackrel{\textbf{E}}{\textbf{H}}
         \end{array}}} $. 
 For the description of the image formation process the interest lies on the excitation strengths of the two propagative fundamental modes in the aperture. These excitation strengths can be calculated by projecting the modes of the aperture onto the investigated field $ t_{x,y}(x,y)=\braket{\textbf{T}_{1,2}^{-}|\tiny{\begin{array}{c}
                  \stackrel{\textbf{E}}{\textbf{H}}
                 \end{array}}} $.
                   A detected image thus is described using Eqn.~\ref{bild0} by
                    \begin{eqnarray}
                                            \mathcal{I}(x,y)=&&\left |\int_{\mathcal{R}^2} [\textbf{e}_{1}^{-}\small(\tilde{x}-x,\tilde{y}-y)\times\textbf{H}(\tilde{x},\tilde{y})-\textbf{E}(\tilde{x},\tilde{y})\times\textbf{h}_{1}^{-}(\tilde{x}-x,\tilde{y}-y)]\,\textbf{n}_z\,d\tilde{x}\,d\tilde{y}\right |^2+ \label{bildges2}\\
                    &&\left |\int_{\mathcal{R}^2} [\textbf{e}_{2}^{-}\small(\tilde{x}-x,\tilde{y}-y)\times\textbf{H}(\tilde{x},\tilde{y})-\textbf{E}(\tilde{x},\tilde{y})\times\textbf{h}_{2}^{-}(\tilde{x}-x,\tilde{y}-y)]\,\textbf{n}_z\,d\tilde{x}\,d\tilde{y} \right |^2 \nonumber,
                           \end{eqnarray}
    where $ \textbf{E}(\tilde{x},\tilde{y}) $ and $ \textbf{H}(\tilde{x},\tilde{y}) $ are the investigated field components at the position of the tip and $ \textbf{e}^{-}_{1,2}(\tilde{x},\tilde{y}) $, $ \textbf{h}^{-}_{1,2}(\tilde{x},\tilde{y}) $ are the mode profiles of the two associated backward propagating mainly $ x $- and $  y $- polarized  modes in the aperture and $ (x,y)^{T} $ defines the position of the tip. 
     This result is in close analogy  to the more general ones presented in Ref.~\cite{Porto2000}. There, the eigenmodal-fields in our case are substituted with the reciprocal fields obtained by evaluating two scenarios fulfilling the requirements of reciprocity.
  
 
 \end{document}